\documentstyle[12pt]{article}

\catcode`\@=11
\long\def\@makefntext#1{
\protect\noindent \hbox to 3.2pt {\hskip-.9pt  
$^{{\ninerm\@thefnmark}}$\hfil}#1\hfill}		

\def\@makefnmark{\hbox to 0pt{$^{\@thefnmark}$\hss}}  
	
\def\ps@myheadings{\let\@mkboth\@gobbletwo
\def\@oddhead{\hbox{}
\rightmark\hfil\ninerm\thepage}   
\def\@oddfoot{}\def\@evenhead{\ninerm\thepage\hfil
\leftmark\hbox{}}\def\@evenfoot{}
\def\sectionmark##1{}\def\subsectionmark##1{}}

\renewcommand{\thefootnote}{\fnsymbol{footnote}}

\newcounter{sectionc}\newcounter{subsectionc}\newcounter{subsubsectionc}
\renewcommand{\section}[1] {\vspace*{0.6cm}\addtocounter{sectionc}{1} 
\setcounter{subsectionc}{0}\setcounter{subsubsectionc}{0}\noindent 
	{\normalsize\bf\thesectionc. #1}\par\vspace*{0.4cm}}
\renewcommand{\subsection}[1] {\vspace*{0.6cm}\addtocounter{subsectionc}{1} 
	\setcounter{subsubsectionc}{0}\noindent 
	{\normalsize\it\thesectionc.\thesubsectionc. #1}\par\vspace*{0.4cm}}
\renewcommand{\subsubsection}[1]
{\vspace*{0.6cm}\addtocounter{subsubsectionc}{1}
	\noindent {\normalsize\rm\thesectionc.\thesubsectionc.\thesubsubsectionc. 
	#1}\par\vspace*{0.4cm}}

\newcounter{appendixc}
\newcounter{subappendixc}[appendixc]
\newcounter{subsubappendixc}[subappendixc]

\renewcommand{\appendix}[1] {\vspace*{0.6cm}
        \refstepcounter{appendixc}
        \setcounter{figure}{0}
        \setcounter{table}{0}
        \setcounter{equation}{0}
        \renewcommand{\thefigure}{\Alph{appendixc}.\arabic{figure}}
        \renewcommand{\thetable}{\Alph{appendixc}.\arabic{table}}
        \renewcommand{\theappendixc}{\Alph{appendixc}}
        \renewcommand{\theequation}{\Alph{appendixc}.\arabic{equation}}
        \noindent{\bf Appendix \theappendixc #1}\par\vspace*{0.4cm}}

\def\abstracts#1{{
	\centering{\begin{minipage}{12.2truecm}\footnotesize\baselineskip=12pt\noindent
	\centerline{\footnotesize ABSTRACT}\vspace*{0.3cm}
	\parindent=0pt #1
	\end{minipage}}\par}} 

\renewenvironment{thebibliography}[1]
	{\begin{list}{\arabic{enumi}.}
	{\usecounter{enumi}\setlength{\parsep}{0pt}
\setlength{\leftmargin 1.25cm}{\rightmargin 0pt}
   	 \setlength{\itemsep}{0pt} \settowidth
	{\labelwidth}{#1.}\sloppy}}{\end{list}}

\newcounter{itemlistc}
\newcounter{romanlistc}
\newcounter{alphlistc}
\newcounter{arabiclistc}

\newcommand{\fcaption}[1]{
        \refstepcounter{figure}
        \setbox\@tempboxa = \hbox{\footnotesize Fig.~\thefigure. #1}
        \ifdim \wd\@tempboxa > 6in
           {\begin{center}
        \parbox{6in}{\footnotesize\baselineskip=12pt Fig.~\thefigure. #1}
            \end{center}}
        \else
             {\begin{center}
             {\footnotesize Fig.~\thefigure. #1}
              \end{center}}
        \fi}

\newcommand{\tcaption}[1]{
        \refstepcounter{table}
        \setbox\@tempboxa = \hbox{\footnotesize Table~\thetable. #1}
        \ifdim \wd\@tempboxa > 6in
           {\begin{center}
        \parbox{6in}{\footnotesize\baselineskip=12pt Table~\thetable. #1}
            \end{center}}
        \else
             {\begin{center}
             {\footnotesize Table~\thetable. #1}
              \end{center}}
        \fi}

\def\@citex[#1]#2{\if@filesw\immediate\write\@auxout
	{\string\citation{#2}}\fi
\def\@citea{}\@cite{\@for\@citeb:=#2\do
	{\@citea\def\@citea{,}\@ifundefined
	{b@\@citeb}{{\bf ?}\@warning
	{Citation `\@citeb' on page \thepage \space undefined}}
	{\csname b@\@citeb\endcsname}}}{#1}}

\newif\if@cghi
\def\cite{\@cghitrue\@ifnextchar [{\@tempswatrue
	\@citex}{\@tempswafalse\@citex[]}}
\def\citelow{\@cghifalse\@ifnextchar [{\@tempswatrue
	\@citex}{\@tempswafalse\@citex[]}}
\def\@cite#1#2{{$\null^{#1}$\if@tempswa\typeout
	{IJCGA warning: optional citation argument 
	ignored: `#2'} \fi}}

 1
 1
 1

\font\ninerm=cmr9

\input psfig

\newcommand{\alt}{\mathrel{\raisebox{-.6ex}{$\stackrel{\textstyle<}{\sim}$}}}
\newcommand{\agt}{\mathrel{\raisebox{-.6ex}{$\stackrel{\textstyle>}{\sim}$}}}
\def\lsim{\alt}
\def\gsim{\agt}

\def\eg{{\it e.g.}}

\def\stop{\wt t}

\def\mstop{m_{\stop}}

\def\hsm{h_{\rm SM}}
\def\mhsm{m_{\hsm}}
\def\hl{h^0}
\def\hh{H^0}
\def\ha{A^0}
\def\hp{H^+}
\def\hm{H^-}
\def\hpm{H^{\pm}}
\def\mhl{m_{\hl}}
\def\mhh{m_{\hh}}
\def\mha{m_{\ha}}

\def\mhpm{m_{\hpm}}
\def\tanb{\tan\beta}

\def\mt{m_t}
\def\mb{m_b}
\def\mz{m_Z}
\def\mw{m_W}

\def\h{h}
\def\mh{m_{\h}}

\def\MPL #1 #2 #3 {Mod.~Phys.~Lett.~{\bf#1},\  #2 (#3)}
\def\NPB #1 #2 #3 {Nucl.~Phys.~{\bf#1},\  #2 (#3)}
\def\PLB #1 #2 #3 {Phys.~Lett.~{\bf#1},\  #2 (#3)}
\def\PR #1 #2 #3 {Phys.~Rep.~{\bf#1},\ #2 (#3)}
\def\PRD #1 #2 #3 {Phys.~Rev.~{\bf#1},\  #2 (#3)}
\def\PRL #1 #2 #3 {Phys.~Rev.~Lett.~{\bf#1},\  #2 (#3)}
\def\RMP #1 #2 #3 {Rev.~Mod.~Phys.~{\bf#1},\  #2 (#3)}
\def\ZP #1 #2 #3 {Z.~Phys.~{\bf#1},\  #2 (#3)}
\def\IJMP #1 #2 #3 {Int.~J.~Mod.~Phys.~{\bf#1},\  #2 (#3)}

\def\taup{\tau^+}
\def\taum{\tau^-}

\def\hpm{H^{\pm}}

\def\call{{\cal L}}

\def\tauptaum{\tau^+\tau^-}

\def\ltot{L_{\rm tot}}
\def\taup{\tau^+}
\def\taum{\tau^-}

\def\br{BF}
\def\tauptaum{\tau^+\tau^-}

\def\gam{\gamma}
\def\sigrts{\sigma_{\tiny\rts}^{}}

\def\sighbar{\overline \sigma_{\h}}

\def\sighhbar{\overline \sigma_{\hh}}
\def\sighabar{\overline \sigma_{\ha}}
\def\anti{\overline}
\def\epem{e^+e^-}
\def\zstar{Z^\star}
\def\wstar{W^\star}

\def\mupmum{\mu^+\mu^-}

\def\rts{\sqrt s}

\def\eg{{\it e.g.}}

\def\anti{\overline}

\def\mw{m_W}
\def\mz{m_Z}
\def\h{h}
\def\mh{m_{\h}}
\def\gamh{\Gamma_{\h}^{\rm tot}}
\def\hsm{h_{SM}}
\def\mhsm{m_{\hsm}}
\def\gamhsm{\Gamma_{\hsm}^{\rm tot}}
\def\tanb{\tan\beta}
\def\hl{h^0}
\def\mhl{m_{\hl}}

\def\ha{A^0}
\def\mha{m_{\ha}}
\def\gamha{\Gamma_{\ha}^{\rm tot}}
\def\hh{H^0}
\def\mhh{m_{\hh}}
\def\gamhh{\Gamma_{\hh}^{\rm tot}}
\def\fbi{~{\rm fb}^{-1}}

\def\mev{~{\rm MeV}}
\def\gev{~{\rm GeV}}
\def\tev{~{\rm TeV}}
\def\stop{\widetilde t}
\def\mstop{m_{\stop}}
\def\mt{m_t}
\def\mb{m_b}

\begin{document}
\noindent{\bf UCD-97-4 \hfill
February, 1997}
\vskip.2in
\centerline{\normalsize\bf HIGGS PHYSICS AT A MUON COLLIDER~\footnote{To
appear in the proceedings of the Ringberg workshop on ``The Higgs Puzzle'',
ed. B. Kniehl, Dec. 8-13, 1996.}}
\baselineskip=15pt
\centerline{\footnotesize JOHN F. GUNION}
\baselineskip=13pt
\centerline{\footnotesize\it Davis Institute for High Energy Physics,
Department
of Physics, University of California at Davis}
\baselineskip=12pt
\centerline{\footnotesize\it Davis, CA 95616, USA}
\vspace*{0.9cm}
\abstracts{A brief review of muon-collider $s$-channel Higgs physics 
in the Standard
Model and its minimal supersymmetric extension is presented.}
 
\normalsize\baselineskip=15pt
\setcounter{footnote}{0}
\renewcommand{\thefootnote}{\alph{footnote}}
\section{Introduction}

One of the most important tasks in high energy physics in the next decades
will be to determine if Higgs boson(s) exist, and to measure the
properties of any that are found. Only then will a full understanding
of the Higgs sector and its role in the complete underlying
theory be possible.
Here we outline the very substantial
and, in some areas, unmatched capabilities of a muon collider 
for these tasks.~\cite{bbgh} We restrict our discussions
to the Standard Model (SM) with a single neutral 
Higgs boson, $\hsm$, and to the minimal
supersymmetric extension of the SM (MSSM) with a light SM-like
Higgs boson, $\hl$,
and two heavier neutral Higgs scalars, $\hh$ and $\ha$.

Although muon colliders have only received substantial attention in
the last few years, so far it seems~\cite{muonreports} that there are
``no barriers'' to either of the following two designs:
\begin{itemize}
\item $\rts\lsim 500\gev$, 
$\call\sim {\rm few} \times 10^{33}$cm$^{-2}$s$^{-1}$,
$\ltot=50\fbi$/yr;
\item $\rts\lsim 4\tev$, $\call\sim {\rm few} \times 10^{35}$
cm$^{-2}$s$^{-1}$, $\ltot=200\fbi$/yr.
\end{itemize}
A $\mupmum$ collider has some natural advantages 
as compared to an $\epem$ collider, including some
that are crucial for Higgs boson physics:
\begin{itemize}
\item there is essentially no beamstrahlung; 
\item there is substantially reduced bremsstrahlung;~\footnote{Still,
bremsstrahlung depletes the central Gaussian peak in $\rts$
by $\sim 40\%$ --- this effect is included in the computations.~\cite{bbgh}}
\item there is no final focus problem (storage rings are used to build up
the effective instantaneous luminosity);
\item as required for detailed Higgs studies in the $s$-channel,
excellent beam energy resolution of $R=0.01\%$ and the ability 
to set $E_{\rm beam}$ to 1 part in $10^6$ using spin
precession techniques are both possible if the necessary technology is
built into the machine;
\item $\rts_{\rm max}> 500\gev$ can probably be reached more easily; and
\item there is hope that the cost might be lower.
\end{itemize}
The negatives regarding a muon collider include:
\begin{itemize}
\item the design is immature, and five years of research and development
projects are needed before a full-fledged proposal would be possible ---
in particular, cooling tests are required to see if multistage 
cooling will be sufficiently efficient;
\item the exact nature of the detector backgrounds, and how to manage them,
is still under investigation --- certainly the detector will be more
expensive due to higher shielding and segmentation requirements;
\item significant polarization probably implies significant loss in $\call$
 --- for $s$-channel Higgs production, 
$S/\sqrt B\propto \sqrt{\ltot}\sqrt{(1+P^2)^2/(1-P^2)}$
implies that
$[P=0,\ltot]$ is equivalent to $[P=0.8,\ltot/10]$;
\item it is not possible to have a $\gam\gam$ collider facility.
\end{itemize}

\begin{figure}[h]
\leavevmode
\begin{center}
\centerline{\psfig{file=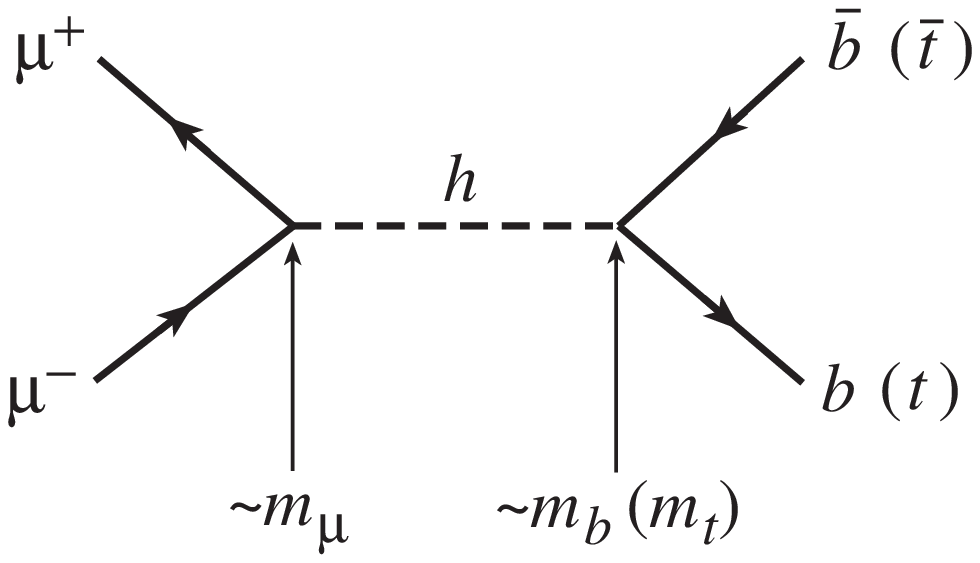,width=5cm}}
\end{center}
\fcaption{Feynman diagram for $s$-channel production of a Higgs boson.}
\label{schanfig}
\end{figure}

\begin{figure}[h]
\leavevmode
\begin{center}
\centerline{\psfig{file=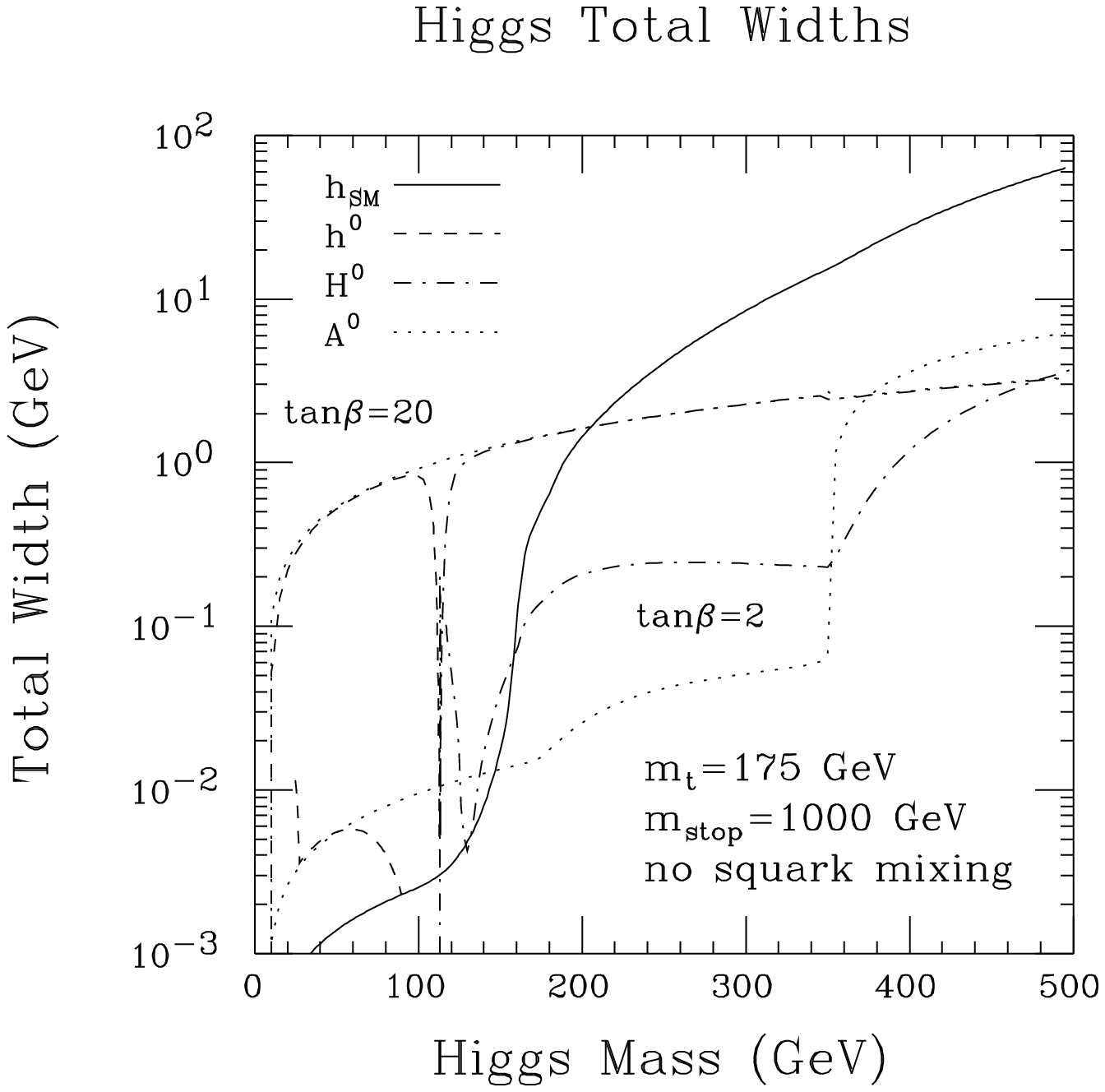,width=3.25in}}
\end{center}
\fcaption{
Total width versus mass of the SM and MSSM Higgs bosons
for $\mt=175\gev$.
In the case of the MSSM, we have plotted results for
$\tan\beta =2$ and 20, taking $\mstop=1\tev$ and
including two-loop/RGE-improved Higgs mass corrections and
neglecting squark mixing; SUSY decay channels are assumed to be absent.}
\label{hwidths}
\end{figure}

A muon collider can explore Higgs physics in two modes. 
Away from the $s$-channel Higgs pole, \eg\ at
the maximum available energy $\rts_{\rm max}$, a $\mupmum$ collider
can discovery and study Higgs bosons in exactly 
the same ways as an $\epem$ collider
with the same $\rts$ and $\call$ (barring
unmanageable detector backgrounds at the muon collider).
On the $s$-channel pole at $\rts\sim \mh$, 
$\mupmum\to \h$ collisions (see Fig.~\ref{schanfig}) imply
unique capabilities for:
\begin{itemize}
\item precision studies of $(b\anti b\h)^2:(WW\h)^2:(ZZ\h)^2$ coupling-squared
 ratios and of $\gamh$ for a SM-like Higgs (either $\hsm$ or $\hl$)
with $\mh\lsim 2\mw$; and
\item discovery and study
of the heavier $\hh,\ha$ of the MSSM up to $\rts_{\rm max}$.
\end{itemize}
For $s$-channel Higgs physics,
the size of the $s$-channel cross section, $\sighbar$, is crucial.
We obtain $\sighbar$ 
by convoluting the standard Breit-Wigner shape for the Higgs
with a Gaussian distribution of width 
$\sigrts$ centered at $\rts=\mh$; $\sighbar$ is given by 
$\sighbar\sim 4\pi\mh^{-2}\br(\h\to\mupmum)$ if $\sigrts\ll\gamh$
and by
$\sighbar\sim2\pi^2\mh^{-2}\Gamma(\h\to\mupmum)/(\sqrt{2\pi}\sigrts)$
if $\sigrts\gg\gamh$. To get near maximal $\sighbar$ and
to have sensitivity to $\gamh$ via scanning in $\rts$ it
is important that $\sigrts$ be no larger than $2-3\times\gamh$.
Very small $\gamh$ is not uncommon.
Fig.~\ref{hwidths} shows that $\gamh< 1-10\mev$ is typical of the $\hsm$
for $\mhsm\lsim 140\gev$ and of {\it all} 
the MSSM Higgs bosons if $\tanb\lsim 2$ and $\mha\lsim 2\mw$.
Using the parameterization 
$\sigrts\simeq 7\mev\left({R\over 0.01\%}\right)\left({\rts\over 
100\gev}\right)$ for $\sigrts$ in terms of the beam energy resolution, $R$,
we see that very excellent resolution $R\sim0.01\%$ is required
for $\sigrts< 2-3\times\gamh$ in the above cases.

A final important note is that high $\call$ is
needed at all $\rts$ values where a Higgs boson 
is discovered or might exist.  This will possibly imply that several
final storage rings (which fortunately are relatively cheap),
designed to maintain near-optimal $\call$ over a span of $\rts$ values, 
will need to be constructed

\section{{\boldmath $s$}-channel studies of a Standard-Model-like Higgs boson}

Most probably, one would first
discover the $\h$ at the LHC (e.g. in the $\h\to\gam\gam$ or $\h\to 4\ell$
discovery modes), 
or at the NLC  (e.g. in the $\zstar\to Z\h$ production mode)
and then set up the muon collider for running at $\rts\sim\mh$. 
At the LHC or NLC, $\mh$ will be quite well determined. 
For $\mh\lsim 2\mw$ one finds:~\cite{snowmass96}
$\Delta\mh\sim 100\mev\left({600\fbi\over L}\right)^{1/2}$ at the LHC, 
using $\h\to\gam\gam,4\ell$ decays and reconstructing the
resonance peak in $m_{\gam\gam},m_{4\ell}$;
$\Delta\mh\sim 100\mev\left({50\fbi\over L}\right)^{1/2}$ for NLC $\rts=500\gev$
running and reconstructing the $m_{b\anti b}$ resonance peak
in the $Z\h$ ($\h\to b\anti b$) mode; and $\Delta\mh< 100\mev$ at the
NLC, assuming $L=50\fbi$ devoted to a $Z\h$ threshold study
at $\rts=\mz+\mh+0.5\gev$. If there is no LHC or NLC, one would accumulate
$L\sim 1\fbi$ at $\rts=500\gev$ at the muon collider
in order to observe $\mupmum\to Z\h$ and
determine $\mh$ to within $\Delta\mh\sim 1\gev$, and then turn to $\rts\sim\mh$
running.

For a SM-like Higgs with $\mh\gsim 2\mw$, 
$\gamh$ is large (as a result
of $\h\to WW,ZZ$ decays), see Fig.~\ref{hwidths}, 
$\sighbar\propto\br(\h\to\mupmum)$ is tiny,
and $\mupmum\to\h$ will not be useful.~\cite{bbgh}
But, if $\mh\lsim 2\mw$ then $\gamh$ is very small,
$\sigrts\sim 2-3\times\gamh$ is typical
for $R=0.01\%$, $\sighbar\propto\Gamma(\h\to\mupmum)/\sigrts$ 
will  be much larger,  and the $\mupmum$ collider
becomes a Higgs factory. There are several reasons to suppose
that the SM-like Higgs will have mass $\lsim 2\mw$:
a) precision electroweak measurements currently
favor (but only weakly) a relatively light SM-like Higgs boson;
b) light Higgs masses are preferred in the SM 
if the SM couplings are required to remain perturbative up to the GUT scale;
and c) in the minimal supersymmetric model (MSSM), the light SM-like Higgs has 
$\mhl\lsim 130\gev$.~\cite{dpfreport}

If a SM-like $\h$ with $\mh\leq 2\mw$ has been discovered, 
the FMC final ring would
be optimized for $\rts\sim \mh$ and a scan of the region
of size $2\Delta\mh$ in the vicinity of the known $\mh$ 
would be performed in order to pin down $\mh$ more
precisely (roughly within $\pm \sigrts/2$).~\footnote{Details of
the strategy
for this scan~\cite{bbgh} will not be discussed here;
the scan is most efficient for small $R$ (as desirable in any case)
in which case $\rts$ must be reset rapidly with high precision.}~
In the  ``typical case'' of $\mh\sim 110\gev$, $\sigrts\sim 8\mev$
(for $R\sim 0.01\%$) and
$2\Delta\mh\sim 200\mev$ (see above), we would require $\gsim 25$ scan points
with $\sim 0.01\fbi$ per scan point (the luminosity for a $5\sigma$
signal if $\rts=\mh$), corresponding to a total luminosity
of $\ltot\sim 0.25\fbi$.
(One must reset $\rts$ with $\Delta\rts<\sigrts$ hourly.)
Centering on $\rts\simeq\mh$ does not require much luminosity      
for such Higgs masses. The worst mass is $\mh\simeq\mz$.
For the $R=0.01\%$ value of $\sigrts\sim 6.5\mev$,
$\ltot\sim 25\fbi$ would be needed to center on $\rts\simeq \mh$
by scanning. A smaller $\Delta\mh$, such as could be obtained
using ``super'' tracking or other improvements
at the NLC,~\cite{snowmass96} would be very helpful for avoiding wasting
luminosity on the  centering process in this case.

Once we have determined $\mh$ to within $\sim \pm\sigrts/2$, we would 
perform a three point scan of the Higgs resonance peak employing
luminosity distributed as follows:
$L_1$ at $\rts\simeq\mh$; $2.5L_1$ at
$\rts\simeq \mh+2\sigrts$; $2.5L_1$ at
$\rts\simeq \mh-2\sigrts$.
The total luminosity for the scan is then $L=6L_1$.  
This scan would simultaneously yield measurements of $\gamh$
and of the $\mupmum\to\h\to b\anti b$, $W\wstar$ and
$Z\zstar$ rates.  Assuming that $L=200\fbi$ is devoted over a period
of several years to this three point scan (presumably beginning with
$\rts\simeq\mh$), one obtains accuracies for
the event rates as given in Table~\ref{fmcsigbrerrors} and
errors for $\gamh$ and coupling-squared ratios as given in
Table~\ref{fmcerrors}.

\begin{table}[h]
\tcaption{Errors for
$\sigma(\mupmum\to\hsm)\br(\hsm\to b\anti b, W\wstar, Z\zstar)$,
for $R=0.01\%$ and $L_{\rm scan}=200\fbi$ 
(equivalent to $L_{\protect\rts=\mh}=50\fbi$).}
\footnotesize
\begin{center}
\small
\begin{tabular}{|c|c|c|c|c|c|}
\hline
 Quantity & \multicolumn{5}{c|}{Errors} \\
\hline
\hline
{$\bf\mhsm$}{\bf (GeV)} & {\bf 80} & {\bf 90} & {\bf 100} & {\bf 110} & {\bf 120} \\
\hline
$\sigma(\mu\mu\to\hsm)\br( b\anti b) $ & 
$\pm 0.2\%$ & $\pm 1.6\%$ & $\pm 0.4\%$ & $\pm 0.3\%$ & $\pm 0.3\%$ \\
\hline
$\sigma(\mu\mu\to\hsm)\br( W\wstar) $ &
$-$ & $-$ & $\pm 3.5\%$ & $\pm 1.5\%$ & $\pm 0.9\%$ \\
\hline
$\sigma(\mu\mu\to\hsm)\br( Z\zstar) $ &
$-$ & $-$ & $-$ & $\pm 34\%$ & $\pm 6.2\%$ \\
\hline
\hline
{$\bf\mhsm$}{\bf (GeV)} & {\bf 130} & {\bf 140} & {\bf 150} & {\bf 160} & {\bf 170} \\
\hline
$\sigma(\mu\mu\to\hsm)\br( b\anti b) $ & 
$\pm 0.3\%$ & $\pm 0.5\%$ & $\pm 1.1\%$ & $\pm 59\%$ & $-$ \\
\hline
$\sigma(\mu\mu\to\hsm)\br( W\wstar) $ &
$\pm 0.7\%$ & $\pm 0.5\%$ & $\pm 0.5\%$ & $\pm 1.1\%$ & $\pm 9.4\%$ \\
\hline
$\sigma(\mu\mu\to\hsm)\br( Z\zstar) $ &
$\pm 2.8\%$ & $\pm 2.0\%$ & $\pm 2.1\%$ & $\pm 22\%$ & $\pm 34\%$ \\
\hline
\hline
{$\bf\mhsm$}{\bf (GeV)} & {\bf 180} & {\bf 190} & {\bf 200} & {\bf 210} & {\bf 220} \\
\hline
$\sigma(\mu\mu\to\hsm)\br( W\wstar) $ &
 $\pm 18\%$ & $\pm 38\%$ & $\pm 58\%$ & $\pm 79\%$ & $-$ \\
\hline
$\sigma(\mu\mu\to\hsm)\br( Z\zstar) $ &
$\pm 25\%$ & $\pm 27\%$ & $\pm 35\%$ & $\pm 45\%$ & $\pm 56\%$ \\
\hline
\end{tabular}
\end{center}
\label{fmcsigbrerrors}
\end{table}

\begin{table}[h]
\tcaption{Errors for 
coupling-squared ratios and $\gamhsm$ for
$s$-channel Higgs production at the FMC, assuming $L=6L_1=200\fbi$
total scan luminosity and $R=0.01\%$.}
\footnotesize
\begin{center}
\begin{tabular}{|c|c|c|c|c|}
\hline
 Quantity & \multicolumn{4}{c|}{Errors} \\
\hline
\hline
{$\bf\mhsm$}{\bf (GeV)} & {\bf 80} & {\bf $\mz$} & {\bf 100} & {\bf 110} \\
\hline
$(W\wstar\hsm)^2/(b\anti b\hsm)^2$ & $-$ & $-$ & $\pm 3.5\%$ & $\pm 1.6\%$ \\
\hline
$(Z\zstar\hsm)^2/(b\anti b\hsm)^2$ & $-$ & $-$ & $-$ & $\pm 34\%$ \\
\hline
$(Z\zstar\hsm)^2/(W\wstar\hsm)^2$ & $-$ & $-$ & $-$ & $\pm 34\%$ \\
\hline
 $\gamhsm$ & $\pm 2.6\%$ & $\pm 32\%$ & $\pm 8.3\%$ & 
  $\pm 4.2\%$ \\
\hline
\hline
{$\bf\mhsm$}{\bf (GeV)} & {\bf 120} & {\bf 130} & {\bf 140} & {\bf 150} \\
\hline
$(W\wstar\hsm)^2/(b\anti b\hsm)^2$ & 
 $\pm 1\%$ & $\pm 0.7\%$ & $\pm 0.7\%$ & $\pm 1\%$ \\
\hline
$(Z\zstar\hsm)^2/(b\anti b\hsm)^2$ & 
 $\pm 6\%$ & $\pm 3\%$ & $\pm 2\%$ & $\pm 2\%$ \\
\hline
$(Z\zstar\hsm)^2/(W\wstar\hsm)^2$ & 
 $\pm 6\%$ & $\pm 3\%$ & $\pm 2\%$ & $\pm 2\%$ \\
\hline
 $\gamhsm$ & $\pm 3.6\%$ & $\pm 3.6\%$ & $\pm 4.1\%$ &
  $\pm 6.5\%$ \\
\hline
\end{tabular}
\end{center}
\label{fmcerrors}
\end{table}
Very good errors in the $c\anti c$ and $\tauptaum$ channels
might also possible, depending on the detector.
For example, $c\anti c$ isolation requires
topological tagging in which one must
distinguish primary, secondary, and 
tertiary vertices for a $b$-jet vs. primary and
secondary only for a $c$-jet. The ability to do so
depends on how close to the beam the first layer 
of the vertex detector can be placed.

What do the errors of Table~\ref{fmcerrors} 
mean in terms of our ability to discriminate between the SM $\hsm$
and the MSSM $\hl$? Plots~\cite{dpfreport,snowmass96}
of the $W\wstar/b\anti b$ coupling-squared
ratio as computed in the MSSM divided by that computed in the SM
show: a) there is almost no dependence of the ratio 
upon the squark mixing scenario; 
b) the ratio is essentially independent of $\tanb$ in the
allowed portion of the standard $(\mha,\tanb)$ parameter space;
and c) roughly $-50\%$ ($-20\%$) deviations from the SM result
are predicted for $\mha\sim 250\gev$ ($\mha\sim 400\gev$).
By the time the muon collider is in operation, theoretical systematic
errors (primarily from uncertainty in the running mass, $\mb(\mh)$)
in the ratio should be in the $\pm 5\%$ to $\pm 10\%$ range.
Table~\ref{fmcerrors} shows that statistical errors will be much smaller
over most of the relevant $\mh$ range.  Thus, the $W\wstar/b\anti b$
event rate ratio will distinguish between the SM and the MSSM
at the $\geq 2\sigma$ level for $\mha\lsim 400\gev$. For $\mha< 400\gev$,
a rough determination of $\mha$ will be possible. (This will be important
for the $\hh,\ha$ discussion in the next section.) 
Similar results apply for the $Z\zstar/b\anti b$ ratio.  

In contrast to the event rate ratios, deviations in $\gamh$ 
in the MSSM vs. the SM as a function of $(\mha,\tanb)$ are very dependent
upon the squark mixing scenario, the amount of SUSY decays present,
and so forth; large deviations in $\gamh$ 
from SM expectations are the rule, but do not pin down either $\mha$
or $\tanb$. However, if one~\cite{snowmass96}
uses the very accurate direct determination
of $\gamh$ from the FMC scan in combination with other measurements
performed with $L=200\fbi$ and $\rts=500\gev$ at the NLC~\footnote{The same
measurements are also possible at an FMC, but we use NLC notation
in what follows.}~
a large variety of very important coupling-squared magnitudes can be extracted.
The ratio of the MSSM prediction to the SM prediction for a squared coupling
is always very squark-mixing independent and gives new opportunities
for determining $\mha$.
As an example, there are four ways to determine
$\Gamma(\h\to\mupmum)$ by combining NLC and FMC data:~\footnote{Note
that since $\sigrts\sim 2-3\times\gamh$, 
the measured $\mupmum\to\h\to X$ rate is more or less proportional
to $\Gamma(\h\to\mupmum)/\sigrts$ so that $\Gamma(\h\to\mupmum)$
can be computed given the known $\sigrts$; small corrections
from the influence of $\gamh$ can be made using the measured $\gamh$.}~
$\Gamma(\h\to\mupmum)=$

\begin{tabular}{ll}
 1)~ ${[\Gamma(\h\to\mupmum)\br(\h\to
b\anti b)]_{\rm FMC}\over \br(\h\to b\anti b)_{\rm NLC}}$;&
 2)~ ${[\Gamma(\h\to\mupmum)\br(\h\to
W\wstar)]_{\rm FMC}\over\br(\h\to W\wstar)_{\rm NLC}}$;\\
 3)~ ${[\Gamma(\h\to\mupmum)\br(\h\to
Z\zstar)\gamhsm]_{\rm FMC}\over\Gamma(\h\to Z\zstar)_{\rm NLC}}$;&
 4)~ ${[\Gamma(\h\to\mupmum)\br(\h\to
W\wstar)\gamhsm]_{\rm FMC}\over\Gamma(\h\to W\wstar)_{\rm NLC}}$.\\
\end{tabular}

\noindent
In the above, $\gamh|_{\rm FMC}$ refers to the scan measurement at the FMC.
Resulting errors for $\h=\hsm$ are tabulated in Table~\ref{nlcfmcerrors},
labelled $(\mupmum\hsm)^2|_{\rm NLC+FMC}$.  The predicted
$(\mupmum\hsm)^2|_{\rm MSSM}/(\mupmum\hsm)^2|_{\rm SM}$ coupling-squared ratio
is independent of $\tanb$: one finds
values of 1.5, 1.2, $1.15$ for $\mha=300$, 475, $600\gev$, respectively (for
theoretically allowed $\tanb$ values at the given $\mha$).
Since there is
no systematic error due to uncertainty in the muon mass
and since experimental systematics should be well below 10\%, 
the above numbers imply that the expected $\lsim \pm 5\%$ statistical
error probes out to very high $\mha$.

\begin{table}[h]
\tcaption{Errors for combining NLC ($L=200\fbi$) data, 
the NLC + LHC ($L=600\fbi$,
ATLAS+CMS) $\br(\hsm\to\gam\gam)$ determination, 
$\gam\gam$ collider ($L=50\fbi$) data and FMC $s$-channel ($L=200\fbi$) data.}
\footnotesize
\begin{center}
\begin{tabular}{|c|c|c|c|c|}
\hline
 Quantity & \multicolumn{4}{c|}{Errors} \\
\hline
\hline
{$\bf\mhsm$}{\bf (GeV)} & {\bf 80} & {\bf 100} & {\bf 110} & {\bf 120} \\
\hline
 $(b\anti b\hsm)^2|_{\rm NLC+FMC} $ & $\pm6\%$ & $\pm 9\%$ & $\pm 7\%$ &
  $\pm6\%$ \\
\hline
 $(c\anti c\hsm)^2|_{\rm NLC+FMC} $ & $\pm9\%$ & $\pm 11\%$ & $\pm 10\%$ &
  $\pm9\%$ \\
\hline
 $(\mupmum\hsm)^2|_{\rm NLC+FMC}$ & 
$\pm 5\%$ & $\pm 5\%$ & $\pm 4\%$ & $\pm 4\%$ \\
\hline
 $(\gam\gam\hsm)^2|_{\rm FMC}$ & $\pm 15\%$ & $\pm 16\%$ & $\pm 14\%$ &
 $\pm 13\%$ \\
\hline
 $(\gam\gam\hsm)^2|_{\rm NLC+FMC}$ & $\pm 9\%$ & $\pm 10\%$ & $\pm 9\%$ &
 $\pm 9\%$ \\
\hline
\hline
{$\bf\mhsm$}{\bf (GeV)} & {\bf 130} & {\bf 140} & {\bf 150} & {\bf 170} \\
\hline
 $(b\anti b\hsm)^2|_{\rm NLC+FMC}$ & $\pm7\%$ & $\pm7\%$ & $\pm10\%$ &
 $\pm23\%$ \\
\hline
 $(c\anti c\hsm)^2|_{\rm NLC+FMC} $ & $\pm10\%$ & \multicolumn{3}{c|}{$?$} \\
\hline
 $(\mupmum\hsm)^2|_{\rm NLC+FMC}$ & 
$\pm 3\%$ & $\pm 3\%$ & $\pm 4\%$ &  $\pm 10\%$ \\
\hline
 $(W\wstar\hsm)^2|_{\rm FMC}$ & $\pm 16\%$ & $\pm 9\%$ & $\pm 9\%$ &
 $-$ \\
\hline
 $(W\wstar\hsm)^2|_{\rm NLC+FMC}$ & $\pm 5\%$ & $\pm 4\%$ & $\pm 6\%$ &
 $\pm 10\%$ \\
\hline
 $(\gam\gam\hsm)^2|_{\rm FMC}$ & $\pm 14\%$ & $\pm 18\%$ & $\pm 36\%$ & $-$ \\
\hline
 $(\gam\gam\hsm)^2|_{\rm NLC+FMC}$ & 
    $\pm 10\%$ & $\pm 13\%$ & $\pm 23\%$ & $-$ \\
\hline
\end{tabular}
\end{center}
\label{nlcfmcerrors}
\end{table}

Table~\ref{nlcfmcerrors} gives a number of other quantities that are
determined with remarkable precision by combining $\rts=500\gev$ NLC data
and $s$-channel FMC data
[and, in the case of $(\gam\gam\hsm)^2$, including the (NLC+LHC) determination
of $\br(\hsm\to\gam\gam)$].~\cite{snowmass96}  The advantages
of having high $L$ data from all three machines are enormous. 
In particular, 
if there is a SM-like Higgs bosons with $\mh\lsim 2\mw$, 
it would be much more beneficial to have 
both an $\epem$ collider operating at full energy, $\rts\sim 500\gev$,
{\it and} a $\mupmum$  collider operating at $\rts\sim \mh$, as opposed
to two NLC's.

\section{The Heavy MSSM Higgs Bosons}

Colliders other than the FMC offer various mechanisms
to directly search for the $\ha,\hh$, but have significant limitations:
\begin{itemize}
\item There are regions in $(\mha,\tanb)$ parameter space at moderate
$\tan \beta$, $\mha\gsim 200\gev$ in which the $\hh,\ha$ cannot be detected
at the LHC.
\item At the NLC one can use the mode $\epem\to \zstar\to \hh\ha$,
but it is limited to $\mhh\sim \mha\lsim \sqrt{s}/2$.
\item A $\gamma \gamma$ collider could probe heavy Higgs up to masses of
$\mhh\sim \mha\sim 0.8\rts$, but this would quite likely require
$L> 100\fbi$, especially if the Higgs bosons are at the upper
end of the  $\gamma \gamma$ collider energy spectrum.~\cite{ghgamgam}
\end{itemize}
In contrast, there is an excellent chance of being able to detect
the $\hh,\ha$ at a $\mupmum$ collider provided only that $\mha$ is smaller
than the maximal $\rts$ available. This could prove to be very important
given that GUT MSSM models usually predict $\mha\gsim 200\gev$.

A detailed study of $s$-channel production
of the $\hh,\ha$ has been made.~\cite{bbgh} 
The signals become viable when $\tanb>1$
(as favored by GUT models) since the $\mupmum\hh$ and $\mupmum\ha$
couplings are proportional to $\tanb$. In particular, 
even though $\gamhh,\gamha$ are big (see Fig.~\ref{hwidths}) at high $\tanb$, 
due to large $b\anti b$ decay widths, $\br(\hh,\ha\to\mupmum)$
approaches a constant value that is large enough to imply
substantial $\sighhbar,\sighabar$. The optimal strategy 
for $\hh,\ha$ detection and study depends upon the circumstances.
First, it could be that the $\hh$ and/or $\ha$ will already have been
discovered at the LHC.  
With $L=300\fbi$ (ATLAS+CMS) of integrated           
luminosity, this would be the case if $\tanb\lsim 3$ or $\tanb$ 
is above an $\mha$-dependent lower bound (\eg\ $\tanb\gsim 10$ for $\mha\sim
400\gev$).~\footnote{For 
$\tanb\lsim 3$, one makes use of modes such as $\hh\to\hl\hl\to
b\anti b \gam\gam$ and $\hh \to ZZ^{(*)}\to 4\ell$, when $\mhh\lsim 2\mt$,
or $\hh,\ha\to t\anti t$, when $\mhh,\mha\gsim 2\mt$.  At high $\tanb$,
the enhanced production rates for $b\anti b \hh,b\anti b\ha$ with
$\hh,\ha\to\taup\taum$ are employed.}~
Even if the $\hh,\ha$ have not been detected,
strong constraints on $\mha$ are possible if precision measurements
of the properties of the $\hl$ (such as the $b\anti b/W\wstar$
and $c\anti c/b\anti b$ event rate ratios and the $(\mupmum\hl)^2$
coupling-squared, as discussed earlier) are made via $s$-channel production
at the FMC or in $\rts=500\gev$ running at the NLC~\cite{snowmass96} or
by combining these two types of data.
By limiting the $\rts$ scan for the $\hh$ and $\ha$
in the $s$-channel to the $\mha\sim\mhh$ 
mass region preferred by $\hl$ measurements,
we would greatly reduce the luminosity needed to
find the $\ha$ and $\hh$ via an $s$-channel scan as compared to that required
if $\mha$ is not constrained.

With such pre-knowledge
of $\mha$, it will be possible to detect and perform detailed
studies of the $\hh,\ha$ for all $\tanb\geq 1$
provided only that $\mha\lsim \rts_{\rm max}$.~\footnote{We
assume that a final ring optimized for maximal luminosity at $\rts\sim \mha$
would be constructed.}~ If $\tanb\lsim 3$ and $\mhh,\mha\lsim 2\mt$, then
excellent resolution, $R\sim 0.01\%$, will be necessary for detection
since the $\ha$ and $\hh$ are relatively narrow (see Fig.~\ref{hwidths}).
For higher $\tanb$ values, $R\sim 0.1\%$ is adequate
for $\hh,\ha$ detection, but $R\sim 0.01\%$ would be required in order
to separate the rather overlapping $\hh$ and $\ha$ peaks (as a function
of $\rts$) from one another.

Even without pre-knowledge of $\mha$, 
there would be an excellent chance for discovery of the $\ha,\hh$
Higgs bosons in the $s$-channel at a $\mupmum$ collider if they
have not already been observed at the LHC.
This is because non-observation at the LHC implies 
that $\tanb\gsim 3$ while it is precisely for
$\tanb\gsim 2.5-3$ that detection of the $\ha,\hh$ is possible~\cite{bbgh}
in the mass range from 200 to 500 GeV via an $s$-channel scan in 
$\mupmum$ collisions. (The lower $\tanb$ reach given
assumes that $\ltot=200\fbi$ is devoted to the scan.
A detailed strategy as to how
much luminosity to devote to different $\rts$ scan settings
in the $200-500\gev$ range must be employed.~\cite{bbgh})
That the LHC and the FMC are complementary in this
respect is a very crucial point. Together, the LHC and FMC
guarantee discovery of the $\ha,\hh$ after 3 to 4 years of
high luminosity operation each, provided $\mha\lsim 500\gev$.
Once $\mha,\mhh$ are known, very precise measurements of some of
the crucial properties of the $\hh,\ha$ 
(including a scan determination of their
total widths) become possible.~\cite{bbgh}

In the event that the NLC has not been constructed, it could be
that the first mode of operation of the FMC would be to optimize
for and accumulate luminosity at, say, $\rts=500\gev$.
In this case, there is still a high probability
for detecting the $\hh,\ha$ if they have not been
observed at the LHC. First, if $\mha\sim\mhh\lsim \rts/2\sim 250\gev$ then
$\mupmum\to \hh\ha$ (and $\hp\hm$) pair production will be observed. Second,
although reduced in magnitude compared to an electron
collider, there is a long low-energy bremsstrahlung tail 
at a muon collider that provides a
self-scan over the full range of $\rts$ values below 
the nominal operating energy.
Observation of $\ha,\hh$ $s$-channel peaks in the $b\anti b$ mass 
($m_{b\anti b}$) distribution
created by this bremsstrahlung tail may be possible.
The region of the $(\mha,\tanb)$ parameter space plane for which
a peak is observable depends strongly on the $m_{b\anti b}$
resolution. For excellent $m_{b\anti b}$
resolution of order $\pm 5\gev$ and integrated luminosity
of $L=200\fbi$ at $\rts=500\gev$, 
the $\ha,\hh$ peak(s) are observable for $\tanb\gsim 4-5$
if $500\gev\geq\mha\geq 250\gev$.~\footnote{Required $\tanb$ values increase
dramatically as one moves into the $\mha\sim \mz$ zone,
but this region is covered by $\hh\ha$ pair production.}

Finally, if neither the LHC nor a FMC scan of the $\leq 500\gev$ region
has discovered the $\hh,\ha$, but supersymmetric particles and the $\hl$
have been observed, we would believe that the $\hh,\ha$ must exist
but have $\mha\sim\mhh\geq
500\gev$ . Analyses of the SUSY spectrum in the GUT context
and precision $\hl$ studies
might have yielded some prejudice for the probable $\mha$, and
an extension of the FMC energy up to the appropriate $\rts\sim \mha$ 
for $s$-channel discovery of the $\hh,\ha$
could be considered. But,
a machine with much higher $\rts$, such as the earlier-mentioned $\rts=4\tev$,
might be most worthwhile.
It has been shown~\cite{gk} 
that such an energy with appropriately matched luminosity
would allow discovery of $\mupmum\to\ha\hh$ 
and $\hp\hm$ pair production,
via the $b\anti b$ or $t\anti t$ decay channels of the $\hh,\ha$
and $t\anti b,\anti t b$ decay channels of the $\hp,\hm$,
up to masses very close to $\mha\sim \mhh\sim \mhpm \sim
2\tev$, even if SUSY decays of the $\hh,\ha,\hpm$ are substantial. 
(This mass range certainly includes that expected in
any supersymmetric model that provides a solution to the
naturalness and hierarchy problems.) 
Detailed studies of the $\hh,\ha,\hpm$ would be possible 
once they were discovered.~\cite{gk,fengmoroi}

\section{Conclusions and Discussion}

In this report, we have briefly reviewed 
the capabilities of a muon collider to explore Higgs physics
in the Standard Model and its minimal supersymmetric extension
via direct $s$-channel Higgs production. If there is a light ($\mh\lsim 2\mw$)
SM-like Higgs boson, an $s$-channel scan of the Higgs resonance
peak provides direct measurements of a number of its important 
properties (in particular $\gamh$)
that cannot be duplicated using $\epem$ (or $\mupmum$) 
collisions at high $\rts$. Thus, if a light SM-like Higgs boson
has already been observed at the LHC or in early operation at the first NLC,
it would be much more useful to follow the first NLC with an FMC,
rather than a second NLC.
In the context of the minimal supersymmetric extension of the SM,
the FMC takes on added value.  At the FMC, 
an appropriately designed $s$-channel scan will allow $\hh,\ha$ discovery 
in the $200\lsim \mha\lsim 500\gev$ range for all $\tanb\gsim 3$. Thus, if
$\hh\ha$ pair production is not seen in $\rts=500\gev$
running at the NLC (or FMC), implying $\mha\gsim 200-240\gev$,
nor at the LHC, implying $\tanb\gsim 3$ (and below an 
$\mha$-dependent upper limit) if $\mha\gsim 200\gev$,
the $\hh,\ha$ {\it will} be observed at the FMC (provided $\mha\lsim 500\gev$).
Once discovered (at any collider), the FMC $s$-channel 
production mode allows (for any $\tanb>1$)
a detailed study of some of their key properties
and a scan determination of their total widths. As reviewed
elsewhere,~\cite{bbgh,snowmass96,dpfreport,perspectives97} a muon collider will
be at least as valuable if the Higgs sector is still more
exotic than the constrained two-doublet MSSM sector. Particularly noteworthy
is the ability of $\mu^-\mu^-$ collisions to probe the
$\mu^-\mu^-$ coupling of a doubly-charged
Higgs boson~\footnote{Such a Higgs is
very likely to be seen at the LHC if it has mass below 1 TeV.~\cite{glp}}~
(as present in many Higgs triplet models)
down to an extraordinarily small coupling magnitude. Overall,
if there are elementary Higgs bosons, a muon collider will
almost certainly be mandated
purely on the basis of its ability to explore the Higgs sector
via $s$-channel factory-like production of Higgs bosons.

\section{Acknowledgements}
I would like to acknowledge the many contributions
of my collaborators, V. Barger, M. Berger, and T. Han,
to the muon collider Higgs physics results. This work was supported
in part by the Department of Energy and by the Davis Institute for
High Energy Physics.

\end{document}